\documentclass[8pt,twocolumn]{extarticle}

\usepackage{fullpage}
\usepackage{times}
\usepackage{natbib}

\usepackage{graphicx}

\title{A RELM earthquake forecast based on pattern informatics}
\author{
  James R. Holliday$^{1,2}$, Chien-chih Chen$^{3,2}$, Kristy F. Tiampo$^{4}$,\\
  John B. Rundle$^{2,1}$, Donald L. Turcotte$^{5}$, and Andrea Donnelan$^{6}$
}
\date{\small
 $^1$ Department of Physics - University of California, Davis, USA \\
 $^2$ Computational Science and Engineering Center - University of
      California, Davis, USA \\
 $^3$ National Central University - Taiwan \\
 $^4$ Department of Earth Sciences - University of Western Ontario, Canada \\
 $^5$ Geology Department - University of California, Davis, USA \\
 $^6$ NASA Jet Propulsion Laboratory, USA
}

\begin{document}
\maketitle

%------------------------------------------------------------------------------
% Abstract                                                                  ---
%------------------------------------------------------------------------------

\begin{abstract}
We present a RELM forecast of future earthquakes in California that is
primarily based on the pattern informatics (PI) method.  This method
identifies regions that have systematic fluctuations in seismicity,
and it has been demonstrated to be successful.  A PI forecast map
originally published on 19 February 2002 for southern California
successfully forecast the locations of sixteen of eighteen M$>$5
earthquakes during the past three years.  The method has also been
successfully applied to Japan and on a worldwide basis.  An
alternative approach to earthquake forecasting is the relative
intensity (RI) method.  The RI forecast map is based on recent levels
of seismic activity of small earthquakes.  Recent advances in the PI
method show considerable improvement, particularly when compared with
the RI method using relative operating characteristic (ROC) diagrams
for binary forecasts.  The RELM application requires a probability for
each location for a number of magnitude bins over a five year period.
We have therefore constructed a hybrid forecast in which we combine
the PI method with the RI method to compute a map of probabilities for
events occurring at any location, rather than just the most probable
locations.  These probabilities are further converted, using
Gutenberg-Richter scaling laws, to anticipated rates of future
earthquakes that can be evaluated using the RELM test.
\end{abstract}

%------------------------------------------------------------------------------
% Introduction                                                              ---
%------------------------------------------------------------------------------

\section{Introduction}
There have been a wide variety of approaches applied to the
forecasting of earthquakes \citep{Turcotte91, Kanamori03}. These
approaches can be divided into two general classes; the first is based
on empirical observations of precursory changes. Examples include
precursory seismic activity, precursory ground motions, and many
others. The second approach is based on statistical patterns of
seismicity. Neither approach has been able to provide reliable
short-term forecasts (days to months) on a consistent basis.

Although short-term predictions are not available, long-term
seismic-hazard assessments can be made.  A large fraction of all
earthquakes occur in the vicinity of plate boundaries, although some
do occur in plate interiors.  It is also possible to assess the
long-term probability of having an earthquake of a specified magnitude
in a specified region. These assessments are primarily based on the
hypothesis that future earthquakes will occur in regions where past,
typically large, earthquakes have occurred \citep{KossobokovKTM00}.  As
we will discuss, a more promising approach is to begin with the
hypothesis that the rate of occurrence of small earthquakes in a
region can be analyzed to assess the probability of occurrence of much
larger earthquakes.

The RELM forecast described in this paper is primarily based on the
pattern informatics (PI) method \citep{RundleTKM02, TiampoRMK02,
TiampoRMG02, RundleTSKS03}.  This method identifies regions of
strongly correlated fluctuations in seismic activity.  These regions
are the locations where subsequent large earthquakes have been shown
to occur, therefore indicating a strong association with the high
stress preceding the main shock.  The fluctuations in seismicity rate
revealed in a PI map have been found to be related to the preparation
process for large earthquakes.  Seismic quiescence and seismic
activation \citep{BowmanOSSS98, WyssH88}, which are revealed by the PI
map, are examples of such preparation processes.  The PI method
identifies the existence of correlated regions of seismicity in
observational data that precede the main shock by months and years.
The fact that this correlated region locates the aftershocks as well
as main shocks leads us to identify this region of correlated
seismicity with the region of correlated high stress
\citep{TiampoRMGK02, TiampoRMK02, TiampoRMG02}.  Finally, our results
with the PI map indicate that the occurrences of future significant
earthquakes are better forecasted by a change (correlated fluctuation)
in the average seismicity rate rather than with the high seismicity
rate itself.

The PI method does not predict earthquakes, rather it forecasts the
regions (hotspots) where earthquakes are most likely to occur in the
relatively near future (typically five to ten years). The objective is
to reduce the areas of earthquake risk relative to those given by
long-term hazard assessments. The result is a map of areas in a
seismogenic region (hotspots) where earthquakes are likely to occur
during a specified period in the future.  In this paper a PI map is
combined with historic seismicity data to produce a map of
probabilities for future large events.  These probabilities can be
further converted, using Gutenberg-Richter scaling laws, to forecast
rates of occurrence of future earthquakes in specific magnitude
ranges.  This forecast can be evaluated using the RELM model.  In the
following we present details of the PI method and the procedure for
producing a composite forecast map.  A discussion on binary forecasts
and forecast verification techniques is given in the appendix.

%------------------------------------------------------------------------------
% The PI method                                                             ---
%------------------------------------------------------------------------------

\section{The PI method}
Our approach divides the seismogenic region to be studied into a grid
of square boxes, or ``pixels'', whose size is related to the magnitude
of the earthquakes to be forecast. The rates of seismicity in each box
are studied to quantify anomalous behavior. The basic idea is that any
seismicity precursors represent changes, either a local increase or
decrease of seismic activity, so our method identifies the locations
in which these changes are most significant during a predefined change
interval. The subsequent forecast interval is the five year time
window during which the forecast is valid.  The box size is selected
to be consistent with the correlation length associated with
accelerated seismic activity \citep{BowmanOSSS98}, and the minimum
earthquake magnitude considered is the lower limit of sensitivity and
completeness of the network in the region under consideration.  The PI
method as applied to California in this paper is composed of the
following steps:
\begin{enumerate}
\item %-----
The seismically active region is binned into boxes of size
$0.1^\circ$ x $0.1^\circ$ and all events having $M \geq 3.0$ are used.
These boxes are labeled $x_i$.  This is also the box size specified for
the RELM forecast.
\item %-----
The seismicity obtained from the ANSS catalog for each day in each box
is uniformly spread over that box and the eight immediately adjacent
boxes (the Moore neighborhood
\citep{Moore62}).  The resulting smoothed intensities for each box is a
time series.
\item %-----
Only the top 10$\%$ most active boxes are considered.  These are the
boxes with the most $M_c \geq 3.0$ earthquakes during the period $t_0$
= 1 January 1950 to $t_2$ = 1 August 2005.  $M_c$ is the cutoff
magnitude for the analysis.
\item %-----
Each time series is normalized in time by subtracting the temporal
mean and dividing by the temporal standard deviation.
\item %-----
Each time series is then normalized in space for each value of time by
subtracting the spatial mean and dividing by the spatial standard
deviation.
\item %-----
Two intensity maps $I_1(x_i,t_b,t_1)$, $I_2(x_i,t_b,t_2)$ are computed
by averaging all the time series from an initial time, $t_b$ to $t_1$
where $t_0 < t_b < t_1$, and then from $t_b$ to $t_2$.  Here $t_0$ = 1
January 1950, $t_1$ = 1 January 1985, and $t_2$ = 1 August 2005.
\item %-----
The intensity change $\Delta I(x_i,t_b,t_1,t_2) = I_2(x_i,t_b,t_2) -
I_1(x_i,t_b,t_1)$ is computed at each location and absolute value is
taken $|\Delta I(x_i,t_b,t_1,t_2)|$.
\item %-----
The average of $|\Delta I(x_i,t_b,t_1,t_2)|$ over all values of $t_0 <
t_b < t_{max}$ is then computed.% In view of the fact that a
%time scale $\tau = t_2 - t_1$ has been implicitly chosen, the time
%$t_{max}$ is chosen to be $t_{max} = t_1 - \tau$.  This choice also
%gives the averaging time periods in the intervals $t_b$ to $t_1$ and
%$t_b$ to $t_2$ more equal weight, thereby excluding the possibility of
%large fluctuations (main shocks) occurring just prior to $t_1$ that
%may receive too much weight if $t_b$ were integrated from $t_0$ to
%$t_1$.
\item %-----
Finally, the mean squared change $<|\Delta I(x_i,t_b,t_1,t_2)|>^2$ is
computed.
\end{enumerate}
Note that steps (2), (3), (7), and (8) have been modified from the
original, published algorithm \citep{RundleTKM02, TiampoRMK02,
TiampoRMG02, RundleTSKS03}.

Hotspot pixels are defined to be the regions where $\Delta
P_i(t_0,t_1,t_2)$ is larger than some threshold value in the interval
$[0,1]$. In these regions, $P_i(t_0,t_1,t_2)$ is larger than the
average value for all boxes (the background level). Note that since
the intensities are squared in defining probabilities the hotspots may
be due to either increases of seismic activity during the change time
interval (activation) or due to decreases (quiescence). We hypothesize
that earthquakes with magnitudes larger than $M_c+2$ will occur
preferentially in hotspots during the forecast time interval $t_2$ to
$t_3$.  Note that this is a binary forecast: either an earthquake is
forecast to occur or it is forecast not to occur.

%------------------------------------------------------------------------------
% Relative Intensity                                                        ---
%------------------------------------------------------------------------------

\section{Relative intensity}
An alternative approach to earthquake forecasting is to use the rate
of occurrence of small earthquakes in the past.  We refer to this type
of forecast as a relative intensity (RI) forecast.  In such a
forecast, the study region is again tiled with boxes of size
$0.1^\circ \times 0.1^\circ$.  The number of earthquakes with
magnitude $M\ge3.0$ in each box is determined over the time period
from $t_0$ to $t_2$.  The RI score for each box is then computed as
the total number of earthquakes in the box in the time period divided
by the value for the box having the largest value.  In order to create
a binary forecast, a threshold value in the interval $[0,1]$ is
selected.  Large earthquakes having $M\ge5$ are then considered
possible only in boxes having an RI value larger than the
threshold. The physical justification for this type of forecast is
that large earthquakes are considered most likely to occur at sites of
high seismic activity.  In this paper we combine our binary PI
forecast with a continuum RI forecast in order to create our continuum
RELM forecast.

%------------------------------------------------------------------------------
% Binary versus continuum forecasts                                         ---
%------------------------------------------------------------------------------

\section{Binary versus continuum forecasts}
The earthquake forecast make by the PI method is a binary forecast.
An earthquake is forecast to occur in the hotspot regions and not to
occur in the other regions, analogous to the issuance of tornado
warnings.  An extensive methodology has been developed in the
atmospheric sciences for forecast verification.  A standard approach
uses contingency tables and relative operating characteristic (ROC)
diagrams \citep{JolliffeS03}.  An example of binary forecast
construction and verification is presented in the appendix.

The alternative to binary forecasts is a continuum forecast.  The
likelihood of an earthquake throughout the entire region is specified,
analogous to temperature forecasts in the atmospheric sciences.  A
common approach to testing the validity of these forecasts is the
maximum likelihood test.  \citet{KaganJ00} were the first to apply
this test to earthquake forecasts.  The maximum likelihood test is not
appropriate for the verification of binary forecasts because they are
overly sensitive to the least probable events.  For example, consider
two forecasts.  The first perfectly forecasts 99 out of 100 events but
assigns zero probability to the last event.  The second assigns zero
probability to all 100 events.  Under a likelihood test, both
forecasts will have the same skill score of $-\infty$.  Furthermore, a
naive forecast that assigns uniform probability to all possible sites
will always score higher than a forecast that misses only a single
event but is otherwise superior.

%------------------------------------------------------------------------------
% Creating the forecast map                                                 ---
%------------------------------------------------------------------------------

\section{Creating the forecast map}
\begin{figure*}[t!]
\begin{minipage}{\columnwidth}
  \flushleft
  \includegraphics[width=\textwidth]{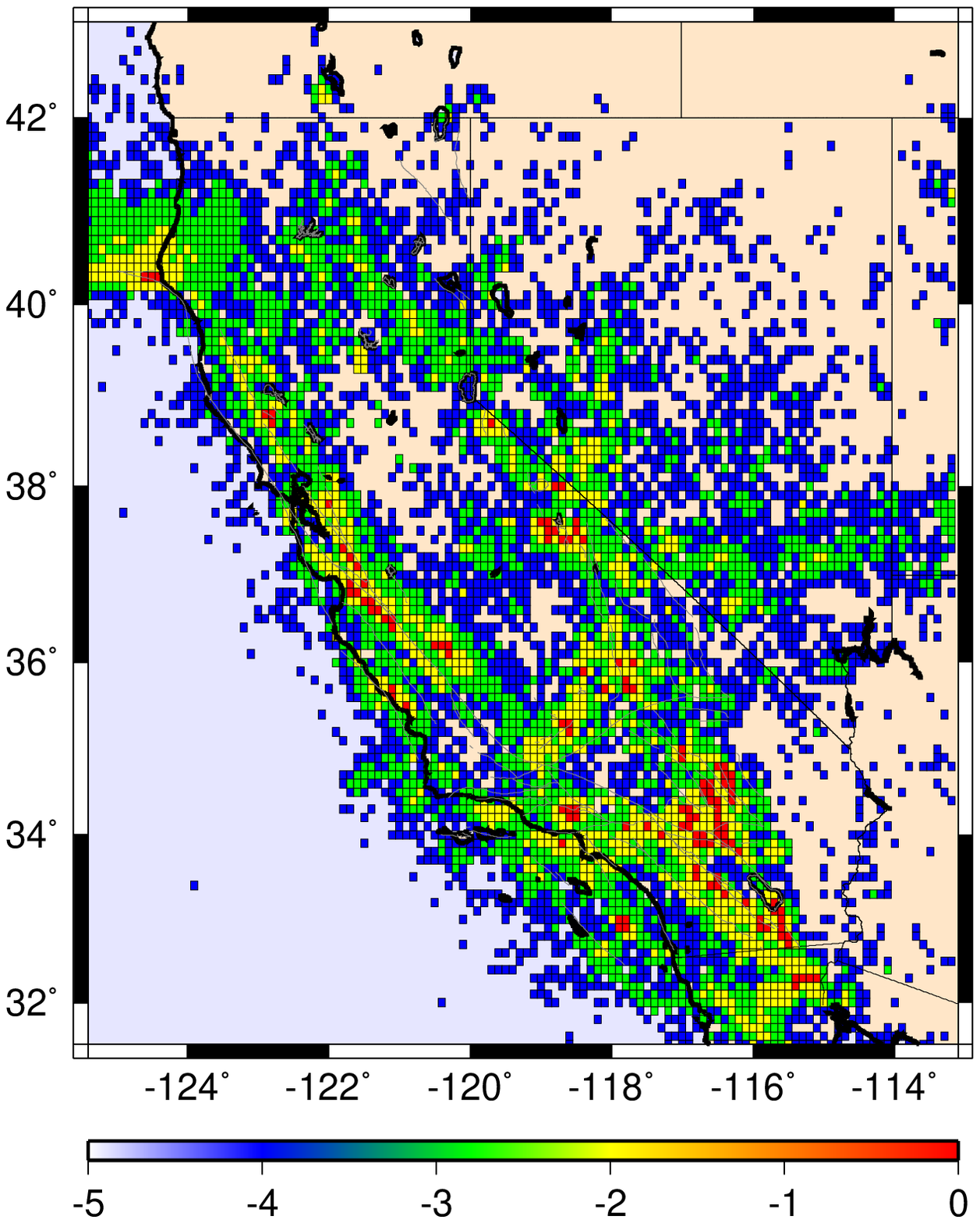} \\ {\bf (A)}
\end{minipage}
\hfill
\begin{minipage}{\columnwidth}
  \flushright
  \includegraphics[width=\textwidth]{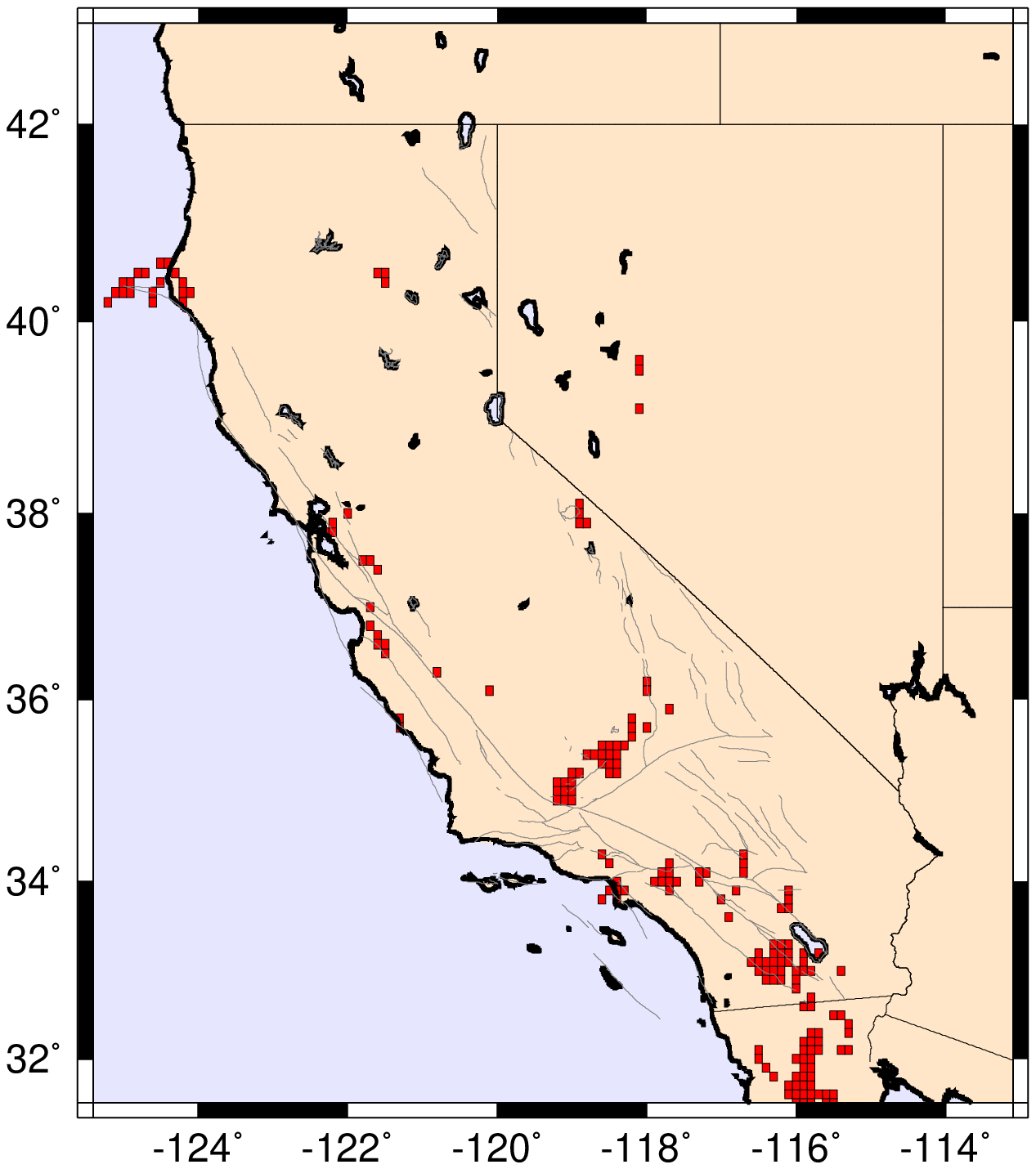} \\ {\bf (B)}
\end{minipage}
\caption{{\bf (A)} Relative intensity (RI) map for all of California
and the surrounding region.  Data from the ANSS on-line catalog for
the years 1950 to 2005 were used.  {\bf (B)} Pattern informatics (PI)
map for the same region and time frame as above.}
\label{f.figure1}
\end{figure*}

\begin{figure*}[t!]
\centering
\includegraphics[width=0.75\textwidth]{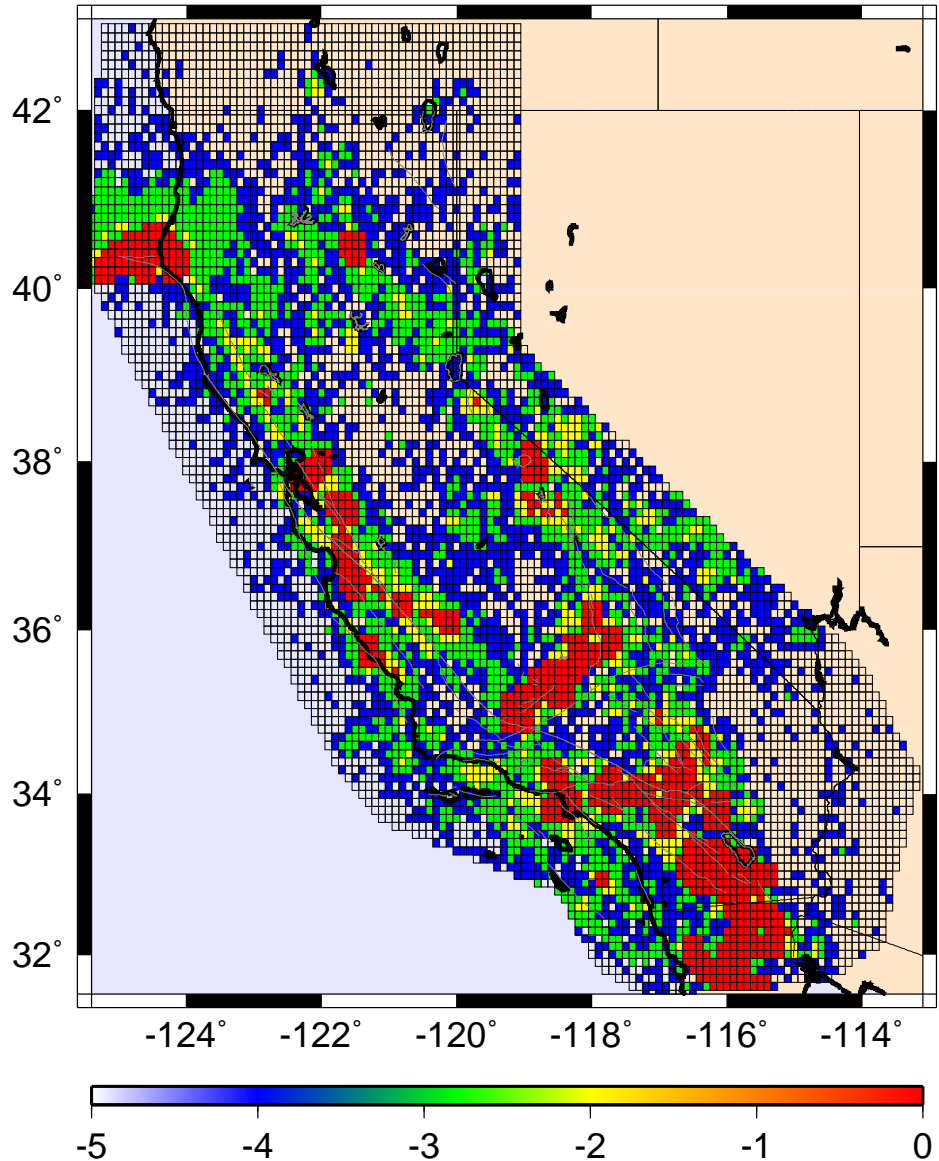}
\caption{Composite forecast map.  The scaled PI and RI maps have
been combined, and boxes outside the testing region have been
discarded.}
\label{f.figure2}
\end{figure*}

The PI method finds regions where earthquakes are most likely to occur
during a future time window.  In order to create a forecast map
suitable for RELM testing, we combined the PI map with the RI map to
create a probability map.  This map is then renormalized to unit
probability and scaled by the total number of $M\ge5$ earthquakes
expected over the future five year period.  The details of this
procedure are as follows:
\begin{enumerate}
\item %-----
We first create a relative intensity map for the entire region to be
considered.  Data was taken from the ANSS on-line catalog for the years
1950 to 2005.  This data was then truncated such that relative values
greater than $10^{-1}$ were set to $10^{-1}$ and non-zero values less
than $10^{-4}$ were set to $10^{-4}$.  Finally, since the RELM
calculations cannot handle zero-rate values, every box with zero
historic seismicity was given a value of $10^{-5}$.  The RI map is
shown in Figure~\ref{f.figure1}A.
\item %-----
We next perform a pattern informatics calculation over the top 10\% of
most active sites in California using the ANSS catalog as input.  For
this calculation, we used $t_0$ = 1 January 1950, $t_1$ = 1 January
1985, and $t_2$ = 1 August 2005.  Since the hotspots are where we
expect future earthquakes to occur, they are given a probability value
of unity.  The PI map is shown in Figure~\ref{f.figure1}B.
\item %-----
We then create a composite probability map by superimposing the PI map
and its Moore neighborhood (the pixel plus its eight adjacent
neighbors) on top of the RI map.  All the hotspot pixels have a
probability of 1, and all other pixels have probabilities that range
from $10^{-5}$ to $10^{-1}$.  The composite map is shown in
Figure~\ref{f.figure2}.
\item %-----
To convert our pixel probabilities to earthquake occurrence
probabilities, we first add up the probabilities in all pixels in the
region and call this sum $N$.  We then normalize this total to the
expected number of $M\geq5.0$ earthquakes during the forecast period.
We estimate four to eight such events per year and assume 30 such
events during a five year period.  In order to do this, we multiply
each pixel probability by $30/N$ to give our RELM forecast.  We then
use Gutenberg-Richter scaling to interpolate these rates into the
appropriate magnitude bins specified by the RELM test.
\end{enumerate}

%\begin{figure}
%\centering
%\begin{minipage}{0.3\columnwidth}
%  {\bf (A)} \vspace{0.25em} \\
%  \includegraphics[width=\columnwidth]{ri_co}
%  \vspace{0.5em}\\
%  {\bf (B)} \vspace{0.2em} \\
%  \includegraphics[width=\columnwidth]{pi_co}
%\end{minipage}
%\hfill
%\begin{minipage}{0.67\columnwidth}
%  \flushright
%  {\bf (C)} \\
%  \includegraphics[width=\columnwidth]{forecast_co}
%\end{minipage}
%\caption{{\bf (A)} Relative intensity (RI) map for all of California
%and the surrounding region.  Data from the ANSS on-line catalog for
%the years 1950 to 2005 were used.  {\bf (B)} Pattern informatics (PI)
%map for the same region and time frame as above.  Only the top 271
%hotspots were retained.  {\bf (C)} Composite forecast map.  The scaled
%PI and RI maps have been combined, and boxes outside the testing
%region have been discarded.}
%\label{f.figure}
%\end{figure}

%------------------------------------------------------------------------------
% Discussion                                                                ---
%------------------------------------------------------------------------------

\section{Discussion}
Ultimately there exists the fundamental question of whether forecasts
of the time and location of future earthquakes can be accurately
made. It is accepted that long term hazard maps of the expected rate
of occurrence of earthquakes are reasonably accurate. But is it
possible to do better? Are there precursory phenomena that will allow
earthquakes to be forecast?

It is actually quite surprising that immediate local precursory
phenomena are not seen. Prior to a volcanic eruption, increases in
regional seismicity and surface movements are generally observed. For
a fault system, the stress gradually increases until it reaches the
frictional strength of the fault and a rupture is initiated. It is
certainly reasonable to hypothesize that the stress increase would
cause increases in background seismicity and aseismic slip. In order
to test this hypothesis the Parkfield Earthquake Prediction Experiment
was initiated in 1985. The expected Parkfield earthquake occurred
beneath the heavily instrumented region on 28 September 2004. No local
precursory changes were observed \citep{Lindh05}.

In the absence of local precursory signals, the next question is
whether broader anomalies develop, and in particular whether there is
anomalous seismic activity. It is this question that is addressed in
this paper. Using a technique that has been successfully applied to
the forecasting of El Ni\~no we have developed a systematic pattern
informatics (PI) approach to the identification of regions of
anomalous seismic activity. Applications of this technique to
California, Japan, and on a world-wide basis have successfully
forecast the location of future earthquakes. We emphasize that this is
not an earthquake prediction. It is a forecast of where future
earthquakes are expected to occur during a future time window of five
to ten years. The objective is to reduce the possible future sites of
earthquakes relative to a long term hazard assessment map.

%------------------------------------------------------------------------------
% Acknowledgments                                                           ---
%------------------------------------------------------------------------------

\section*{Acknowledgments}
This work has been supported by NASA Headquarters under the Earth
System Science Fellowship Grant NGT5 (JRH), by research support from
the National Science Council and the Department of Earth Sciences
(CCC), by an HSERC Discovery grant (KFT), by a grant from the US
Department of Energy, Office of Basic Energy Sciences to the
University of California, Davis DE-FG03-95ER14499 (JRH and JBR), and
through additional funding from NSF grant ATM-0327558 (DLT) and the
National Aeronautics and Space Administration under grants through the
Jet Propulsion Laboratory (AD) to the University of California, Davis.

%------------------------------------------------------------------------------
% Forecast Verification                                                     ---
%------------------------------------------------------------------------------

\section*{Appendix A - Forecast verification}
Along with the RELM model, previous published tests of earthquake
forecasts have emphasized the likelihood test \citep{KaganJ00,
RundleTKM02, TiampoRMK02, HollidayRTKD05b}.  As discussed above, these
tests have the significant disadvantage that they are overly sensitive
to the least probable events.  For this reason, likelihood tests are
subject to unconscious bias.

An extensive review on forecast verification in the atmospheric
sciences has been given by \citet{JolliffeS03}.
The wide variety of approaches that they consider are directly
applicable to earthquake forecasts as well.  We believe that many of
these approaches are better suited to the evaluation of earthquake
forecasts.  The earthquake forecasts considered in this paper can be
viewed as binary forecasts by considering the events (earthquakes) as
being forecast either to occur or not to occur in a given box.  We
consider that there are four possible outcomes for each box, thus two
ways to classify each hotspot, box, and two ways to classify each
non-hotspot, box:

\begin{enumerate}
\item An event occurs in a hotspot box or within the Moore neighborhood
      of the box (the Moore neighborhood is comprised of the eight
      boxes surrounding the forecast box).  This is a success.
\item No event occurs in a white non-hotspot box.  This is also
      a success.
\item No event occurs in a hotspot box or within the Moore neighborhood of
      the hotspot box.  This is a false alarm.
\item An event occurs in a white, non-hotspot box.  This is a failure
      to forecast.
\end{enumerate}

We note that these rules tend to give credit, as successful forecasts,
for events that occur very near hotspot boxes.  We have adopted these
rules in part because the grid of boxes is positioned arbitrarily on
the seismically active region, thus we allow a margin of error of $\pm
1$ box dimension.  In addition, the events we are forecasting are
large enough so that their source dimension approaches, and can even
exceed, the box dimension meaning that an event might have its
epicenter outside a hotspot box, but the rupture might then propagate
into the box.  Other similar rules are possible but we have found that
all such rules basically lead to similar results.

The standard approach to the evaluation of a binary forecast is the
use of a relative operating characteristic (ROC) diagram
\citep{Swets73, Mason03}.  Standard ROC diagrams consider the fraction
of failures-to-predict and the fraction of false alarms.  This method
evaluates the performance of the forecast method relative to random
chance by constructing a plot of the fraction of failures to predict
against the fraction of false alarms for an ensemble of forecasts.
Molchan \citep{Molchan97} has used a modification of this method to evaluate
the success of intermediate term earthquake forecasts.

The binary approach has a long history, over 100 years, in the
verification of tornado forecasts \citep{Mason03}.  These forecasts
take the form of a tornado forecast for a specific location and time
interval, each forecast having a binary set of possible outcomes.  For
example, during a given time window of several hours duration, a
forecast is issued in which a list of counties is given with a
statement that one or more tornadoes will or will not occur.  A
$2\times2$ {\it contingency table\/} is then constructed, the top row
contains the counties in which tornadoes are forecast to occur and the
bottom row contains counties in which tornadoes are forecast to not
occur.  Similarly, the left column represents counties in which
tornadoes were actually observed, and the right column represents
counties in which no tornadoes were observed.

With respect to earthquakes, our forecasts take exactly this form.  A
time window is proposed during which the forecast of large earthquakes
having a magnitude above some minimum threshold is considered valid.
An example might be a forecast of earthquakes larger than $M=5$ during
a period of five or ten years duration.  A map of the seismically
active region is then completely covered (``tiled'') with boxes of two
types: boxes in which the epicenters of at least one large earthquake
are forecast to occur and boxes in which large earthquakes are
forecast to not occur.  In other types of forecasts, large earthquakes
are given some continuous probability of occurrence from 0\% to 100\%
in each box \citep{KaganJ00}.  These forecasts can be converted to the
binary type by the application of a {\it threshold\/} value.  Boxes
having a probability below the threshold are assigned a forecast
rating of {\it non-occurrence\/} during the time window, while boxes
having a probability above the threshold are assigned a forecast
rating of {\it occurrence\/}.  A high threshold value may lead to many
{\it failures to predict\/} (events that occur where no event is
forecast), but few {\it false alarms\/} (an event is forecast at a
location but no event occurs).  The level at which the threshold is
set is then a matter of public policy specified by emergency planners,
representing a balance between the prevalence of failures to predict
and false alarms.

%------------------------------------------------------------------------------
% Binary Earthquake Forecast Verification                                   ---
%------------------------------------------------------------------------------

\section*{Appendix B - Binary earthquake forecast verification}
To illustrate this approach to earthquake forecast verification, we
have constructed two types of retrospective binary forecasts for
California.  The first type of forecast utilizes the PI results
published by Rundle {\it et al.} and Tiampo {\it et al.}
\citep{RundleTKM02, TiampoRMK02} for southern California and adjacent
regions ($32^\circ$ to $38.3^\circ$ N lat, $238^\circ$ to $245^\circ$
E long).  This forecast was constructed for the time period 1 January 2000
to 31 December 2009, but we performed an interim analysis using data
up to the present.  The second type of forecast utilizes the
RI results with the same parameter thresholds.

The first step in our generation of ROC diagrams is the construction
of the $2\times2$ contingency table for the PI and RI forecast maps.
The hotspot boxes in each map represent the forecast locations.  A
hotspot box upon which {\it at least\/} one large future earthquake
during the forecast period occurs is counted as a {\it successful
forecast\/}.  A hotspot box upon which {\it no\/} large future
earthquake occurs during the forecast period is counted as an {\it
unsuccessful forecast\/}, or alternately, a {\it false alarm\/}.  A
white box upon which {\it at least\/} one large future earthquake
during the forecast period occurs is counted as a {\it failure to
forecast\/}.  A white box upon which {\it no\/} large future
earthquake occurs during the forecast period is counted as a {\it
unsuccessful forecast of non-occurrence\/}.

Verification of the PI and RI forecasts proceeds in exactly the same
was as for tornado forecasts.  For a given number of hotspot boxes,
which is controlled by the value of the probability threshold in each
map, the contingency table (see Table~\ref{table.contingency}) is
constructed for both the PI and RI maps.  Values for the table
elements $a$ (Forecast=yes, Observed=yes), $b$ (Forecast=yes,
Observed=no), $c$ (Forecast=no, Observed=yes), and $d$ (Forecast=no,
Observed=no) are obtained for each map.  The fraction of colored
boxes, also called the {\it probability of forecast of occurrence\/},
is $r=(a+b)/N$, where the total number of boxes is $N=a+b+c+d$.  The
{\it hit rate\/} is $H=a/(a+c)$ and is the fraction of large
earthquakes that occur on a hotspot.  The {\it false alarm rate\/} is
$F=b/(b+d)$ and is the fraction of non-observed earthquakes that are
incorrectly forecast.

\begin{table}
\caption{\label{table.contingency}
Contingency tables as a function of false alarm rate.  In
Table~\ref{table.contingency}A, a threshold value was chosen such that
$F \approx 0.005$.  In Table~\ref{table.contingency}B, a threshold value was
chosen such that $F \approx 0.021$.}
{\bf (A)}
\begin{center}
\begin{tabular}{|c|c|c|c|}
\multicolumn{4}{c}{Pattern informatics (PI) forecast} \\ \hline
Forecast & \multicolumn{3}{c|}{Observed} \\ \cline{2-4}
         &  Yes    & No       & Total    \\ \hline
Yes      & (a) 4   & (b) 25   & 29       \\
No       & (c) 13  & (d) 4998 & 5011     \\ \hline
Total    &  17     & 5023     & 5040     \\ \hline
\end{tabular}
\begin{tabular}{|c|c|c|c|}
\multicolumn{4}{c}{Relative intensity (RI) forecast} \\ \hline
Forecast & \multicolumn{3}{c|}{Observed} \\ \cline{2-4}
         &  Yes    & No       & Total    \\ \hline
Yes      & (a) 2   & (b) 27   & 29       \\
No       & (c) 14  & (d) 4997 & 5011     \\ \hline
Total    &  16     & 5024     & 5040     \\ \hline
\end{tabular}
\end{center}
\vspace{2Em}

{\bf (B)}
\begin{center}
\begin{tabular}{|c|c|c|c|}
\multicolumn{4}{c}{Pattern informatics (PI) forecast} \\ \hline
Forecast & \multicolumn{3}{c|}{Observed} \\ \cline{2-4}
         &  Yes    & No       & Total    \\ \hline
Yes      & (a) 23  & (b) 104  & 127      \\
No       & (c) 9   & (d) 4904 & 4913     \\ \hline
Total    &  32     & 5008     & 5040     \\ \hline
\end{tabular}
\begin{tabular}{|c|c|c|c|}
\multicolumn{4}{c}{Relative intensity (RI) forecast} \\ \hline
Forecast & \multicolumn{3}{c|}{Observed} \\ \cline{2-4}
         &  Yes    & No       & Total    \\ \hline
Yes      & (a) 20  & (b) 107  & 127      \\
No       & (c) 10  & (d) 4903 & 4913     \\ \hline
Total    &  30     & 5010     & 5040     \\ \hline
\end{tabular}
\end{center}
\end{table}

\begin{figure}
  \centering
  \includegraphics[angle=270,width=\columnwidth]{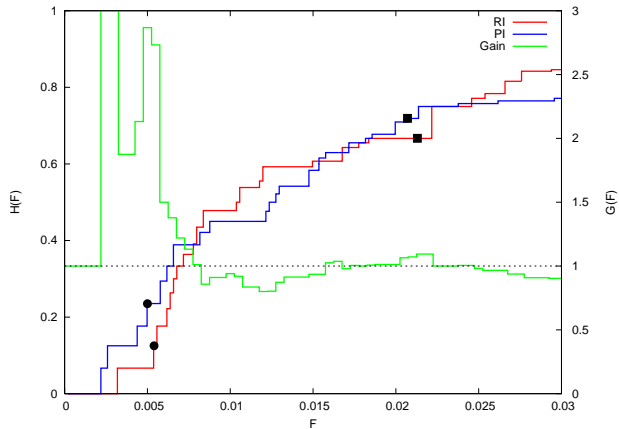}
  \caption{\label{fig.roc}
  Relative operating characteristic (ROC) diagram.  Plot of hit rates,
  $H$, versus false alarm rates, $F$, for the PI forecast and
  RI forecast.  Also shown is the gain ratio defined as
  $H_{PI}(F) / H_{RI}(F)$.  The filled symbols correspond to the
  threshold values used in Table~\ref{table.contingency}, solid
  circles for 29 hotspot boxes and solid squares for 127 hotspot
  boxes.  The horizontal dashed line corresponds to zero gain.}
\end{figure}

\begin{figure}
  \centering
  \includegraphics[angle=270,width=\columnwidth]{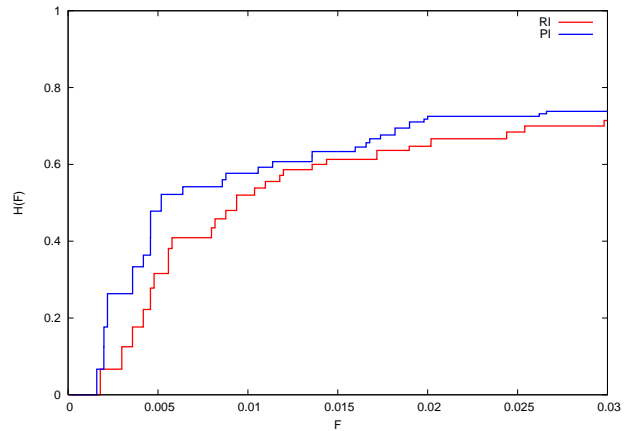}
  \caption{\label{fig.rocnew}
  Relative operating characteristic (ROC) diagram.  Plot of hit rates,
  $H$, versus false alarm rates, $F$, for the RI forecast and
  PI forecast using the modified method.  Note that the PI approach
  outperforms the RI under all circumstances.}
\end{figure}

To analyze the information in the PI and RI maps, the standard
procedure is to consider all possible forecasts together.  These are
obtained by increasing $F$ from 0 (corresponding to no hotspots on the
map) to 1 (all active boxes on the map are identified as hotspots).
The plot of $H$ versus $F$ is the relative operating characteristic
(ROC) diagram.  Varying the threshold value for both the PI and RI
forecasts, we have obtained the values of $H$ and $F$ given in
Figure~\ref{fig.roc}.  The results corresponding to the contingency
tables given in Table~\ref{table.contingency} are given by the filled
symbols.  The forecast with 29 hotspot boxes has $F_{PI} = 0.00498$, $H_{PI} =
0.235$ and $F_{RI} = 0.00537$, $H_{RI} = 0.125$.  The forecast with
127 hotspot boxes has $F_{PI} = 0.0207$, $H_{PI} =
0.719$ and $F_{RI} = 0.0213$, $H_{RI} = 0.666$.  Also shown in
Figure~\ref{fig.roc} is a gain curve defined by the ratio of
$H_{PI}(F)$ to $H_{RI}(F)$.  Gain values greater than unity indicate
better performance using the PI map than using the RI map.  The
horizontal dashed line corresponds to zero gain.  From
Figure~\ref{fig.roc} it can be seen that the PI approach outperforms
(is above) the RI under many circumstances and both outperform a
random map, where $H=F$, by a large margin.  For reference, ROC
diagrams using the modified method discussed in the main text for the
same forecast period are given in Figure~\ref{fig.rocnew}.  Note that
a different input catalog was used for this analysis.  Also note that
in this case, the PI approach outperforms the RI under all
circumstances.

%------------------------------------------------------------------------------
% Bibliography                                                              ---
%------------------------------------------------------------------------------

%%%%%%%%%%%%%%%%%%%%%%%%%%%%%%%%%%%%%%%%%%%%%%%%%%%%%%%%%%%%%%%%%%%%%%%%%%%%%%%
%%% That's all, folks                                                       %%%
%%%%%%%%%%%%%%%%%%%%%%%%%%%%%%%%%%%%%%%%%%%%%%%%%%%%%%%%%%%%%%%%%%%%%%%%%%%%%%%

\end{document}